\documentclass[aps,prc,twocolumn,showpacs,groupedaddress]{revtex4}

\usepackage{graphicx}
\usepackage{epstopdf}
\usepackage{multirow,dcolumn}

\begin{document}

\newcommand{\nnprime}{n,n$^\prime \gamma$}
\newcommand{\ntwon}{n,2n$\gamma$}
\newcommand{\nthreen}{n,3n$\gamma$}
\newcommand{\nxn}{n,xn$\gamma$}
\newcommand{\nx}{n,x$\gamma$}

\newcommand{\natpb}{$^{\textrm{nat}}$Pb}
\newcommand{\natge}{$^{\textrm{nat}}$Ge}
\newcommand{\enrge}{$^{\textrm{enr}}$Ge}
\newcommand{\eigpb}{$^{208}$Pb}
\newcommand{\sevpb}{$^{207}$Pb}
\newcommand{\sixpb}{$^{206}$Pb}
\newcommand{\fivpb}{$^{205}$Pb}
\newcommand{\foupb}{$^{204}$Pb}
\newcommand{\nonubb}  {$0 \nu \beta \beta$}
\newcommand{\twonubb} {$2 \nu \beta \beta$}
\newcommand{\gam}{$\gamma$}
\def\nuc#1#2{${}^{#1}$#2}
\def\mee{$\langle m_{\beta\beta} \rangle$}
\def\mnu{$\langle m_{\nu} \rangle$}
\def\ml{$m_{lightest}$}
\def\gnu{$\langle g_{\nu,\chi}\rangle$}
\def\mmod{$\| \langle m_{\beta\beta} \rangle \|$}
\def\mb{$\langle m_{\beta} \rangle$}
\def\BBz{$0 \nu \beta \beta$}
\def\BBm{$\beta\beta(0\nu,\chi)$}
\def\BBt{$2 \nu \beta \beta$}
\def\nonubb{$0 \nu \beta \beta$}
\def\twonubb{$2 \nu \beta \beta$}
\def\BB{$\beta\beta$}
\def\Mz{$M_{0\nu}$}
\def\Mt{$M_{2\nu}$}
\def\MzG{$M^{GT}_{0\nu}$}           
\def\MzF{$M^{F}_{0\nu}$}                
\def\MtG{$M^{GT}_{2\nu}$}           
\def\MtF{$M^{F}_{2\nu}$}                
\def\Tz{$T^{0\nu}_{1/2}$}
\def\Tt{$T^{2\nu}_{1/2}$}
\def\Tc{$T^{0\nu\,\chi}_{1/2}$}
\def\Rz{$\Gamma_{0\nu}$}            
\def\Rt{$\Gamma_{2\nu}$}            
\def\ms{$\delta m_{\rm sol}^{2}$}
\def\ma{$\delta m_{\rm atm}^{2}$}
\def\ts{$\theta_{\rm sol}$}
\def\ta{$\theta_{\rm atm}$}
\def\tot{$\theta_{13}$}
\def\gpp{$g_{pp}$}                  
\def\qval{$Q_{\beta\beta}$}                 
\def\MJ{{\sc Majorana}}             
\def\DEM{{\sc Demonstrator}}             
\def\be{\begin{equation}}
\def\ee{\end{equation}}
\def\cpRty{counts/ROI/t-y}
\def\onecpRty{1~count/ROI/t-y}
\def\fourcpRty{4~counts/ROI/t-y}
\def\ppc{P-PC}                          
\def\nsc{N-SC}                          


\title{Fast-Neutron Activation of Long-Lived Isotopes in Enriched Ge}

\newcommand{\lanl}{Physics Division, Los Alamos National Laboratory, Los Alamos, NM 87545}
\newcommand{\lansce}{LANSCE Division, Los Alamos National Laboratory, Los Alamos, NM 87545}
\newcommand{\lanlx}{XCP Division, Los Alamos National Laboratory, Los Alamos, NM 87545}
\newcommand{\uw}{Department of Physics, University of Washington, Seattle WA, 98195}

\affiliation{\lanl}
\affiliation{\lansce}
\affiliation{\uw	}

\author{S.R.~Elliott}\email[]{elliotts@lanl.gov}\affiliation{\lanl}
\author{V.E.~Guiseppe}\thanks{Present address: University of South Dakota, Vermillion, SD 57069}\affiliation{\lanl}
\author{R.A.~Johnson}\affiliation{\uw}
\author{B.H.~LaRoque} \affiliation{\lanl}
\author{S.G.~Mashnik}\affiliation{\lanlx}

\date{\today}

\begin{abstract}
We measured the production of \nuc{57}{Co}, \nuc{54}{Mn}, \nuc{68}{Ge}, \nuc{65}{Zn}, and \nuc{60}{Co} in a sample of Ge enriched in isotope 76 due to high-energy neutron interactions. These isotopes, especially \nuc{68}{Ge}, are critical in understanding background in Ge detectors used for double-beta decay experiments. They are produced by cosmogenic-neutron interactions in the detectors while they reside on the Earth's surface. These production rates were measured at neutron energies of a few hundred MeV. We compared the measured production to that predicted by cross-section calculations based on CEM03.02. The cross section calculations over-predict our measurements by approximately a factor of three depending on isotope. We then use the measured cosmic-ray neutron flux, our measurements, and the CEM03.02 cross sections to predict the cosmogenic production rate of these isotopes. The uncertainty in extrapolating the cross section model to higher energies dominates the total uncertainty in the cosmogenic production rate.   
\end{abstract}


\maketitle
 

\section{Introduction}
\label{sec:Intro}

Neutrinoless double-beta decay (\BBz) plays a key role in understanding the neutrino's absolute mass scale and particle-antiparticle nature~\cite{Ell02, Ell04, Bar04, Avi05, eji05, avi08}. If this nuclear decay process exists, one would observe a mono-energetic line originating from a material containing an isotope subject to this decay mode. One such isotope that may undergo this decay is $^{76}$Ge.
Germanium-diode detectors fabricated from material enriched in $^{76}$Ge have established the best half-life
limits and the most restrictive constraints on the effective Majorana mass for the neutrino~\cite{aal02a,bau99}. One analysis~\cite{kla06} of the data in Ref.~\cite{bau99} claims evidence for the decay with a half-life of $2.23^{+0.44}_{-0.31} \times 10^{25}$ y.
Planned Ge-based \BBz\ experiments, \MJ~\cite{ell08,gui08,hen09} and GERDA~\cite{sch05}, will test this claim. Eventually, these future experiments target a sensitivity of $>$10$^{27}$ y or $\sim$1 event/ton-year to explore neutrino mass values near that indicated by the atmospheric neutrino oscillation results. 

The key to these experiments lies in the ability to reduce intrinsic radioactive
background to unprecedented levels and to adequately shield the detectors from external
 sources of radioactivity. Previous experiments' limiting backgrounds have been trace levels of natural decay chain isotopes within the detector and shielding components. The \gam-ray emissions from these isotopes can deposit energy in the Ge detectors producing a continuum, which may overwhelm the potential \BBz\ signal peak at 2039 keV. Great progress has been made in identifying the location and origin of this contamination, and future efforts will substantially reduce this contribution to the background. The background level goal of 1 event/ton-year, however, is an ambitious factor of $\sim$400 improvement over the currently best achieved background level~\cite{bau99}. If the efforts to reduce the natural decay chain isotopes are successful, previously unimportant components of the background must be understood and eliminated. The contribution from long-lived isotopes produced by cosmic-ray neutrons in Ge detectors fabricated from enriched Ge was recognized and described in~\cite{mil92,avi92}. In fact, the dominant background that the  \MJ\ and GERDA experiments will face, without sophisticated analysis cuts, will originate from such isotopes unless mitigation strategies to reduce the activation are successful.
 
 To successfully mitigate the impact of these isotopes requires an understanding of their production. Tables~\ref{tab:NatGeProdRate} and~\ref{tab:EnrGeProdRate} summarize the previous production rate estimates. The two most critical cosmogenic isotopes, \nuc{60}{Co} and especially \nuc{68}{Ge}, show significant variation in the predicted rates: factors of $\approx$2 and $\approx$10 respectively. For  \nuc{60}{Co}, some models predict a higher production rate for enriched Ge (\enrge) than for Ge samples with natural isotopic content (\natge).  Reference~\cite{ceb06} gives a nice summary of previous attempts to calculate the production rates of the problematic isotopes, and provides an estimate of its own. The authors of that report noted that the calculations differ significantly and that measurements would be useful to better understand the rates. Some measurements do exist, but are either  for proton reactions~\cite{hor83,ale95,nor05}, or have a large uncertainty~\cite{avi92}. Since the production rates due to neutrons are much larger than for protons, its important to have neutron reaction measurements. In addition, Barabanov {\em et al.}~\cite{bara06} studied the reduction of cosmogenic activation as it depends on shielding in order to design an optimum transport container. That reference also calculated the rates for cosmic-ray proton reactions and found the production rates for the troublesome isotopes to be about a factor of 10 below that for cosmic-ray neutrons. The work of Mei {\em et al.}~\cite{mei09} noted that much of the large variation in these rates was due to the use of different cosmic ray neutron flux estimates, and that many of the analyses (\cite{mil92, avi92, bal92, bara06, bac08}) used historical flux spectra that are less precise than modern measurements. Although Ref.~\cite{mei09} performed calculations for the production of the isotopes of interest to this paper, the cross sections were calculated using TALYS~\cite{kon04}. This code only predicts cross sections to an energy of 250 MeV, whereas other treatments go to higher energies. Reference~\cite{ceb06} used a modern interpretation of the old neutron flux values~\cite{Zieg98} but not the results of recent measurements~\cite{Gord04}. Furthermore, it used a combination of cross section calculations in order to span the energies necessary for the calculations. Reference~\cite{kla02b} presented numbers for a specific shielding geometry, which is not easy to translate to a raw production rate with the provided data. Hence none of the presently available estimates are sufficient to reliably predict the cosmogenic production rates and new measurements/estimates are required.
 
 We exposed a sample of Ge enriched in isotope 76 to a wide-band neutron beam that resembles the cosmic-ray neutron flux. After exposure we counted the sample in a low-background counting system to observe the \gam\ rays from the decays of the problematic isotopes. From these data we have measured the production rate due to fast neutrons in a Ge sample enriched in isotope 76 that was taken from material used for Ge detector production. With knowledge of the neutron-beam and cosmic-neutron energy spectra, we used these data to provide an estimate of the production rate due to exposure of Ge to cosmic rays. We used a cross-section calculation that spans the energy range of interest and the most recent cosmic-ray neutron flux measurements of which we are aware. This article describes our determination of values for the production rate of these isotopes.
 
 \begin{table}[t]

\caption{A summary of previous estimates of the production of long-lived cosmogenic isotopes in \natge\ for the isotopes studied in this work. The production rates are given in atoms/(kg d). The data in Ref.~\cite{kla02b} was quoted in $\mu$Bq/kg and we converted to units presented.}
\label{tab:NatGeProdRate}
\begin{tabular}{|c|c|c|c|c|c|c|c|c|}

\hline  
Isotope			&  Ref.			& Ref.			         &  Ref.			      &  Ref.		 & Ref.				& Ref.			& Ref.   & Ref.   \\
				&\cite{mil92}   & \cite{avi92}  			  &\cite{avi92}		 & \cite{kla02b} & \cite{bara06}    & \cite{ceb06}  & \cite{bac08} & \cite{mei09}   \\
				&			   & 				(Calc.)   &				(Expt.) & 			 & 				    & 				 &			 & 				   \\
\hline\hline
\nuc{57}{Co}    & 0.5           &  4.4                    &   2.9$\pm$0.4       &     10.2      &                  &   9.7         & 	6.7	& 13.5 \\
\nuc{54}{Mn}    &               &  2.7                    &   3.3$\pm$0.8       &     9.1       &                  &   7.2         & 		& 2.7 \\
\nuc{68}{Ge}    & 26.5          & 29.6                     &   30$\pm$7         &   58.4 	    & 82.8            &   89             &	45.8	&  41.3\\
\nuc{65}{Zn}    & 30.0          &  34.4                    &  38$\pm$6          &  79.0 	        &                 &  77          & 	29.0	 &  37.1 \\
\nuc{60}{Co}    & 4.8           &                          &                     &     6.6      & 2.9             &   4.8          &	2.8	&   2.0 \\
\hline 
\end{tabular}
\end{table} 

 \begin{table}[t]

\caption{A summary of previous estimates of the production of long-live cosmogenic isotopes in \enrge\ for the isotopes studied in this work. For these estimates, the abundance values are 14\% for \nuc{74}{Ge}, 86\% for \nuc{76}{Ge} and zero for the other naturally occurring isotopes. The production rates are given in atoms/(kg d). }
\label{tab:EnrGeProdRate}
\begin{tabular}{|c|c|c|c|c|c|c|c|c|}

\hline  
Isotope			& Ref.			 & Ref.			 &  Ref.         &  Ref.		     	&   Ref. 		& Ref.			& Ref. 			 &  This   \\
				& \cite{mil92}	 &  \cite{avi92} &  \cite{bal92}  &   \cite{bara06}  &\cite{ceb06}   &\cite{bac08}	 & \cite{mei09}     &  Work   \\
\hline\hline
\nuc{57}{Co}   &      0.1 		& 1.0  			&  	1.6		    &  				    &   2.3   		&	2.9			& 6.7  				&   0.7$\pm$0.4 \\
\nuc{54}{Mn}  &  				&  1.4 			&  	2.3		    &  				    &   5.4			&	2.2			& 0.87 				&   2.0$\pm$1.0  \\
\nuc{68}{Ge}    &   1.2	 		& 1.2 			&  	   		    &   5.7			    &   13 			&	7.6			&   7.2				&   2.1$\pm$0.4  \\
\nuc{65}{Zn}    &    6.0   		&  6.4			&  11.0		    & 				    &  24			&	10.4			& 20.0 				&   8.9$\pm$2.5\\
\nuc{60}{Co}   &    3.5			&  				&  				    &   3.3 			    &   6.7			&	2.4			&    1.6 			&   2.5$\pm$1.2 \\
\hline 
\end{tabular}
\end{table}

\section{Experiment}


The sample was exposed to the neutron beam at the Los Alamos Neutron Science Center (LANSCE) Weapons Neutron Research (WNR) facility from Target 4 Flight Path 60 Right (4FP60R) \cite{lis90}. As the broad-spectrum, pulsed neutron beam strikes the Ge target, the outgoing \gam\ rays are detected by the GErmanium Array for Neutron Induced Excitations (GEANIE) spectrometer \cite{bec97}. The corresponding data from the GEANIE spectrometer will be used for (n,n'$\gamma$) analysis that will be presented in a separate publication. The GEANIE sample is located a distance of 20.34 m from the natural tungsten spallation target.

\begin{figure}
\includegraphics[width=8cm]{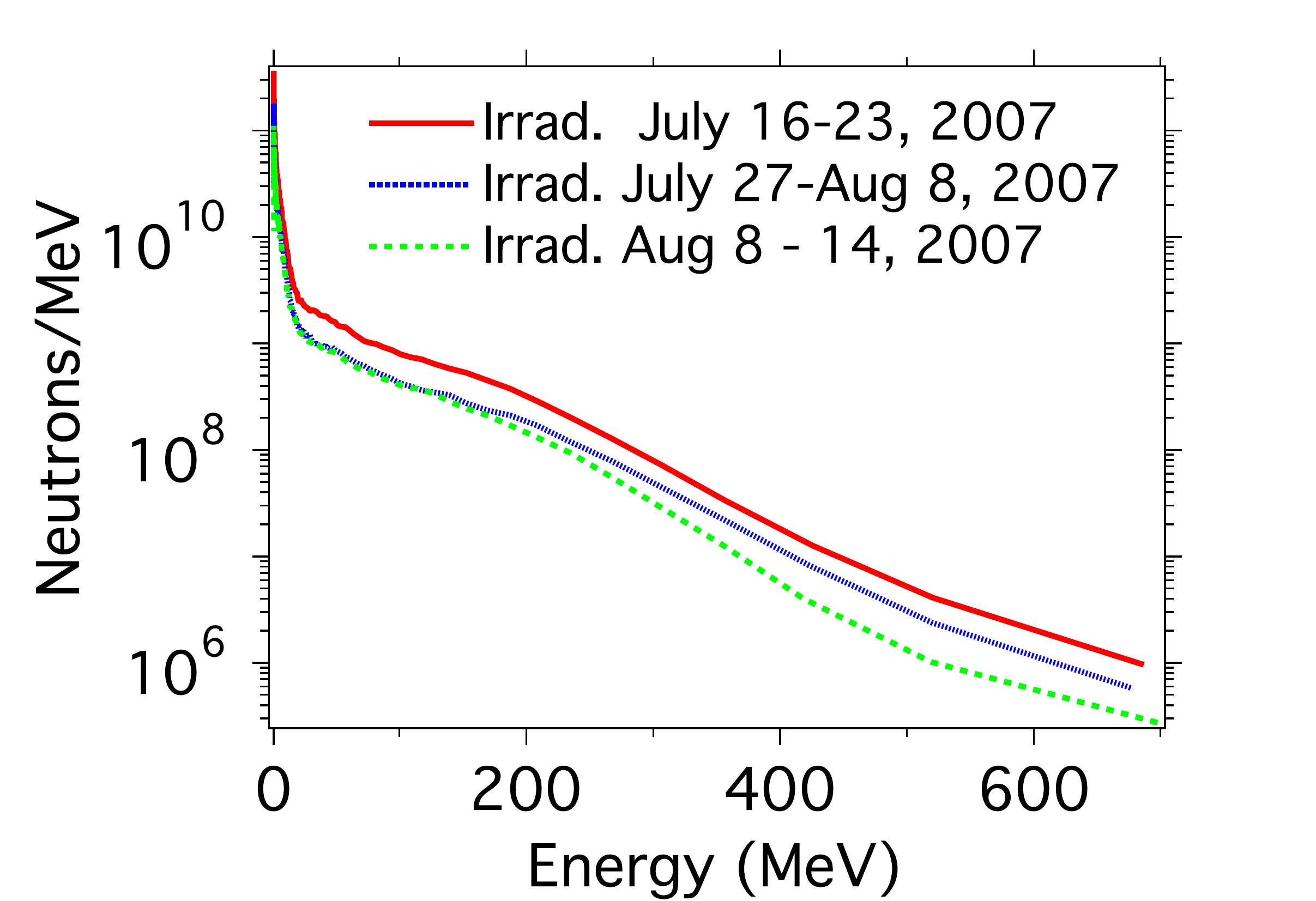}
\caption{The energy spectrum for the neutron beam fluence at 4FP60R for the 3 separate exposure periods. The data are corrected for live time of the fission chamber.}
\label{fig:nFlux}
\end{figure}

The target sample was a 11.13-gm, 22-mm diameter metal enriched Ge (\enrge) powder contained within a plastic enclosure. The isotopic abundances within the sample were measured by time of flight secondary ion mass Spectrometry (ToF SIMS) with the result: \nuc{70}{Ge} $0.77 \pm 0.04$\%, \nuc{72}{Ge} $0.94 \pm 0.05$\%,\nuc{73}{Ge} $0.36 \pm 0.03$\%,\nuc{74}{Ge} $13.81 \pm 0.18$\%,and \nuc{76}{Ge} $84.12 \pm 0.23$\%. The sample was exposed with two separate beam collimations (3/4" and 1/2" collimators) during 3 irradiation periods. For the 3/4"-collimator (1/2"-collimator) run a surface area of 2.85 (1.27) cm$^2$ was exposed to the beam. The 3/4"-collimator exposure was performed between July 16 and Juy 23,  2007 (6.99 d elapsed time). The 1/2"-collimator exposure was performed between July 27 and August 2, 2007 (5.95 d) and then between August 8 and August 14, 2007 (6.19 d). As seen in Fig.~\ref{fig:nFlux}, the energy spectrum of the third exposure was slightly softer than the other 2 exposures at the higher energies. The pulsed neutron beam has the following timing structure. Macropulses, lasting 625 $\mu$s, occur at a rate of 40 Hz. Micropulses are spaced every 1.8 $\mu$s, during which the neutron energy is determined by the time of flight from the micropulse start. An in-beam fission chamber measures the neutron flux with $^{238}$U foils. If the reader wishes to convolve the neutron spectrum with his/her own cross section model, we give a parameterization of the neutron spectrum impinging upon our sample for convenience. The spectrum can be described as:

\begin{eqnarray}
\label{eq:GEANIENFlux}
\Phi(E) & = & (1.325\times10^{-10}) \times  \\
        &  & e^{(4.986\ln E - 3.825\ln^2E+ 0.9159\ln^3E - 0.07402\ln^4E)}  \nonumber 
\end{eqnarray}

\noindent where $\Phi$ is in units of neutrons/MeV and the energy (E) is in MeV.

The powder was stored for an extended period and therefore any radioactivity had decayed to an extremely low level ($<$150 Bq) before counting began. Therefore, this sample was well below any action levels and not subject to any source-handling requirements. The sample was transported to our low-background counting facility underground at the Waste Isolation Pilot Plant (WIPP) near Carlsbad, NM and counted with a Ge detector. The detector was fabricated in 1985 and placed underground at WIPP in 1998. It is an n-type semi-coax design with a height of 41 mm and a diameter of 51 mm. It is contained within an $\approx$1-mm thick Cu cryostat. The shield during these runs consisted of 5 cm of oxygen-free, high-conductivity Cu and 10 cm of Pb. 

\begin{figure}
\includegraphics[width=8cm]{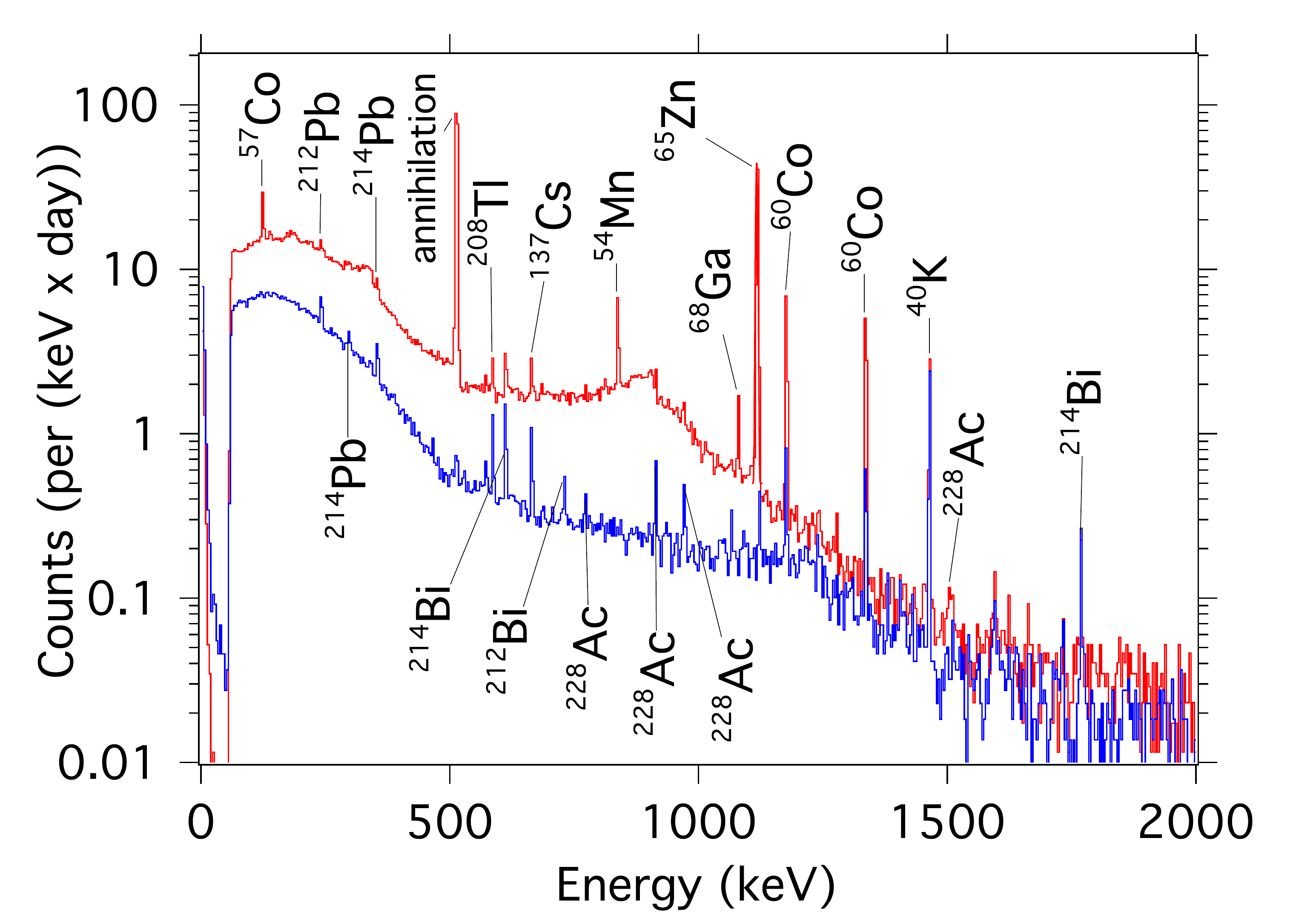}
\caption{The energy spectrum of \gam\ rays from the \enrge\ sample as measured by a Ge detector (upper curve). Also shown is a background spectrum taken with no sample present. (lower curve)}
\label{fig:WIPPnSpectrum}
\end{figure}

The sample was counted over a period of 73.86 days between February 19 and May 4, 2009 with a total live time of 49.02 days. The spectrum is shown in Fig.~\ref{fig:WIPPnSpectrum}. The detector at WIPP has been underground there for over 10 years and therefore any activitives of isotopes of interest to this study, that may have been produced while that detector was exposed to cosmic rays as it resided on the Earth's surface, have long since decayed away. The lone exception to this is a very low level of $^{60}$Co that resides in the Cu cryostat of the Ge detector and the inner layer of the shield that is also Cu. The background spectrum in Fig.~\ref{fig:WIPPnSpectrum} shows that this rate is very small compared to the sample's $^{60}$Co rate. The data presented in Table~\ref{tab:CountResults} includes a subtraction of this background ($\approx$10\%) for the \nuc{60}{Co} count rates.

\section{Measured Production Rates}
The peaks in Fig.~\ref{fig:WIPPnSpectrum} were fit to determine the measured counts ($C$) and the results are given in Table~\ref{tab:CountResults}. The efficiency for counting the \gam\ rays in the geometry used for the sample was determined by using \nuc{57,60}{Co}, \nuc{54}{Mn}, \nuc{22}{Na} and \nuc{137}{Cs} sources. The geometry of these sources is coincidently very similar to the Ge sample and therefore there was no need for simulation to determine the efficiencies. These sources had calibrated activities known to $\pm 1$\%. Since many of the lines of interest come from these isotopes, the measurements were a direct calibration of the efficiencies of interest. For \nuc{65}{Zn} and \nuc{68}{Ga} lines, a \gam-ray efficiency curve was determined from the various lines in these sources in order to interpolate between the measured line energies. Corrections for summing of coincident transitions and annihilation \gam\ rays within the sources were made. The resulting efficiencies and their uncertainty resulting from counting statistics and the interpolation and summing corrections are given in Table~\ref{tab:CountResults}. For the \nuc{60}{Co} calibration, a high event rate resulted in a minor peak distortion that the other data sets did not suffer. We took an additional uncertainty for those peaks to reflect that. We used a random pulser to verify the event-rate dependence of the data-acquisition system dead time.

In addition to the \gam-ray detection efficiency ($\epsilon_{\gamma}$), an efficiency factor must also be included to take into account the decay of the isotope since the end of exposure and during the counting period. This latter efficiency factor ($\epsilon_c$) depends on the half-life of the isotope of interest and is also given in Table~\ref{tab:CountResults}.

With these efficiencies, it is straight-forward to calculate the number of atoms of each isotope that were present at the end of exposure to neutrons (August 14, 2007). The measured number of atoms at the reference date of the end of exposure ($N_0$) is given by:

\begin{equation}
N_0 = \frac{C}{\epsilon_{\gamma} \epsilon_c B},
\end{equation}

\noindent where 

\begin{equation}
\label{eq:epsilonC}
 \epsilon_c = \sum_i({e^{-\lambda T_{start}^i} - e^{-\lambda T_{end}^i} }),
 \end{equation}

\noindent and $T_{end}^i$ ($T_{start}^i$) is the number of days since the end of exposure that the counting stopped (started) for each of the $i$ data runs, $\lambda$ is the decay rate of the isotope in question, B is the fraction of decays that emit the $\gamma$ ray of interest.

The predicted number of atoms {$N^{Pred}_0$ can also be expressed in terms of the reaction rates during the 3 exposure periods:

\begin{equation}
\label{eq:exposure}
N^{Pred}_0 = \sum_i{\frac{k_i}{\lambda}(1 - e^{-\lambda T_{irrad}^i})e^{-\lambda T_{decay}^i}}
\end{equation} 

\noindent where we have corrected for the decay during the exposure and the decay after the exposure until the reference date.
 In Eqn.~\ref{eq:exposure},  $T_{irrad}^i$ is the duration time of irradiation of the $i^{th}$ exposure,  and $T_{decay}^i$ is the time between the end of exposure $i$ and the reference date for $N_0$. $k_i$ is the production rate (atoms/day) during exposure $i$ and is given by:
 
 \begin{eqnarray}
 k_i & = & \frac{n_s}{t_i} \int{F_i(E) \sigma(E) dE} \nonumber \\
      & = & \left( \frac{M_s N_A}{(MW) A t_i} \right)  f_i
 \end{eqnarray}

\noindent where $n_s$ is the number of exposed sample nuclei per unit area in exposure $i$ of duration $t_i$, $F_i(E)$ is the energy ($E$) dependent neutron fluence (number of neutrons per MeV) impinging upon the sample during exposure $i$ and $\sigma(E)$ is the energy dependent cross section calculated for our sample and its measured isotopic abundances. $f_i$ is the fluence and cross section energy-dependent integral for each exposure. The areal density, $n_s$, is related to the total sample mass ($M_s$, 11.13 g) and the sample area ($A$, 3.80 cm$^2$). The beam spot covers most, but not all, of 
the sample geometry and therefore we have ignored any uncertainty associated with non-uniformities in the powdered sample thickness. That is, any such non-uniformities will average to the nominal areal density.


$N^{Pred}_0$ can be written:

\begin{equation}
\label{eq:ProdRate}
N^{Pred}_0  =  \frac{M_s N_A}{(MW) \lambda A}\sum_i{f_i (1 - e^{-\lambda T_{irrad}^i})e^{-\lambda T_{decay}^i}}  
\end{equation}

\noindent where $N_A$ is Avagadro's number, and $MW$ is the molecular weight (75.62 u for our sample).

\begin{table*}[t]

\caption{A summary of $N_0$ for the long-lived isotopes in the \enrge\ sample. The uncertainty quoted for $N_0$ is statistical and arises only from the counting statistics of the measured peaks. The quoted uncertainty for $N^{Pred}_0$ arises from the uncertainty in the 4FP60R neutron fluence. Systematic uncertainties are summarized in Table~\ref{tab:CosmicRateUnc} and discussed in the text.}
\label{tab:CountResults}
\begin{tabular}{|c|c|c|c|c|c|c|c|c|}

\hline  
Isotope		& $\tau_{1/2}$ (days)&  \gam-ray Energy	& $C$			& $\epsilon_{\gamma}$	&  $\epsilon_c$	&  $B$	&  $N_0$							& $N^{Pred}_0$ \\
\hline\hline
\nuc{57}{Co}	&   271.8			&   ~122.1  keV		& 2916$\pm$84	& 0.1663(7)				&  0.0280		& 0.856	&$7.31 \pm 0.21  \times 10^{5}$	& $2.97\pm0.06 \times 10^6$\\
\nuc{57}{Co}	&   271.8			&   ~136.5  keV		& 386$\pm$63		& 0.163(2)				&  0.0280		& 0.107	&$7.89 \pm 1.3   \times 10^{5}$	& $2.97\pm0.06 \times 10^6$\\
\nuc{54}{Mn}	&   312.1			&     ~834.9 keV		&   1084$\pm$43	&  0.0302(3)				&  0.0297		&1.000	&$1.21 \pm 0.05  \times 10^{6}$  & $1.79\pm0.05 \times 10^6$\\
\nuc{68}{Ge}	&   270.8			&   1077.4 keV		& ~198$\pm$18	&  0.0207(5)				&  0.0280		& 0.032	&$1.06 \pm 0.10  \times 10^{7}$	&$2.92\pm0.02 \times 10^7$\\
\nuc{65}{Zn}	&  244.25			&  1115.5 keV		&   8541$\pm$95	& 0.0232(5)				&  0.0262		& 0.506	&$2.77 \pm 0.03  \times 10^{7}$	&$5.03\pm0.06 \times 10^7$\\
\nuc{60}{Co}	&   1923.6			&     1173.2  keV	& 1342$\pm$42	& 0.0200(6)				&  0.0146		& 0.999	&$4.61 \pm 0.14  \times 10^{6}$	&$7.50\pm0.01 \times 10^6$\\
\nuc{60}{Co}	&   1923.6			&  1332.5  keV		& 1176$\pm$39	& 0.0167(3)				&  0.0146		& 1.000	&$4.83 \pm 0.16  \times 10^{6}$	&$7.50\pm0.01 \times 10^6$ \\
\hline 
\end{tabular}
\end{table*} 

A number of systematic effects add to the uncertainty in $N_0$. These include the uncertainty in the nuclear physics parameters, which come from the National Nuclear Data Center. The values for the half-lives and branching ratios are known to high precision and are a negligible contribution to the total uncertainty. The uncertainly in $\epsilon_{\gamma}$ is described above and included in the quoted values. However, each of the sources have an additional uncertainty due to the precision of the known activity of 1\%. The start and stop times of counting and the live time of the counting are known to a small precentage and are negligible contributions to the uncertainty. This is similar for the times associated with the irradiation.  The sample had been stored on the Earth's surface for many years prior to exposure to the beam and then counting at WIPP. Any isotopes produced by cosmic ray neutrons would certainly be at saturation after this extended period. The saturation production rates however are predicted to be near 1 atom/(day kg). Therefore the total count rate for our 11 g sample due to this contribution is several orders of magnitude less than our measured count rate and we ignore this systematic effect.

One can assign an effective energy at which these measurements were made as that for which the theoretical production rate (the product of the CEM cross section and the 4FP60R flux) peaks. Table~\ref{tab:CrossSectionsExtrap} provides the effective energy for each of the isotopes in question. The measured uncertainty in the production rates at these energies are given in Table~\ref{tab:CosmicRateUnc} in the {\em subtotal} column.

\section{Cross Sections}
\label{sec:CrossSection}
If the energy dependence of the neutron fluence over the duration of the exposure had been constant and identical in shape to the cosmic-ray neutron flux, we could have used a simple neutron-flux scaling between our measurements and the cosmic-neutron flux to estimate the cosmogenic production of these isotopes. 
Since the shapes do differ, we used a model for the cross section to adjust for these spectral differences in order to make a prediction regarding the cosmic-ray production rates. For this purpose, we use cross sections for isotope production calculated using Cascade-Exciton Model (CEM03.02~\cite{gud83,mas05,mas08}) as it usually has a better predictive power in comparison with other similar available models (see, e.g.~\cite{Tit08}). 
The CEM formalism permits the calculation of cross sections to the high energies necessary for this work. We assumed that the energy dependence of the cross section is nearly correct but that there may be an overall normalization uncertainty. This procedure then corrects for any normalization uncertainty. Below we describe how we estimate the uncertainty associated with the energy dependence assumption. 

To determine the yield of an isotope, one must consider all feeder isotopes that may decay to the isotope of interest. Therefore to determine the cumulative cross section for \nuc{57}{Co} for example, one must sum the cross sections for \nuc{57}{Co}, \nuc{57}{Ni}, \nuc{57}{Cu} and \nuc{57}{Zn}. Similarly for \nuc{68}{Ge}, one must also consider \nuc{68}{As} and for \nuc{65}{Zn} one must consider \nuc{65}{Zn}, \nuc{65}{Ga}, \nuc{65}{Ge} and \nuc{65}{As}. For the isotopes \nuc{54}{Mn} and \nuc{60}{Co} one only need consider the primary isotopes. Note, that for the case of \nuc{60}{Co}, this is only approximately true. Because of the long half-life of \nuc{60}{Fe}, the contributions of \nuc{60}{Fe}, \nuc{60}{Mn} and \nuc{60}{Cr} are negligible. Using the CEM3.02 code, Tables~\ref{tab:CrossSections70} through \ref{tab:CrossSections76} provide the values of these cumulative cross sections summed over the feeders for the 5 isotopes that comprise Ge. 

\begin{table}[t]
\caption{The calculated cumulative cross sections for production of selected cosmogenic isotopes in \nuc{70}{Ge}. The cross sections are cumulative because they are sums over all isotopes that feed the isotope of interest. }
\label{tab:CrossSections70}
\begin{tabular}{|c|c|c|c|c|c|}

\hline  
Energy            & \multicolumn{5}{c|}{Isotope Production Cross Section (mb)} \\
\cline{2-6}
(MeV)              &\nuc{57}{Co} & \nuc{54}{Mn} &\nuc{68}{Ge}    &\nuc{65}{Zn}    &\nuc{60}{Co}    \\
\hline\hline
10	&0			&0			&0			&0			&0 \\
20	&0			&0			&0			&0			&0 \\
30	&0			&0			&295.6		&16.63		&0 \\
40	&0			&0			&248.0 		&97.52		&0 \\
50	&0			&0			&153.1		&51.05		&0.0022 \\
60	&0			&0			&119.6		&36.86		&0.034 \\
70	&0			&0			&101.6		&55.99		&0.53 \\
80	&0.009		&0.001		&90.53		&86.66		&0.80 \\
90	&0.019		&0.0019		&82.67		&91.14		&0.70 \\
100	&0.16		&0.0037		&75.15		&88.94		&0.80 \\
200	&11.49		&3.06		&50.42		&54.34		&4.78 \\
300	&16.52		&7.72		&43.50		&47.17		&4.34 \\
400	&18.97		&10.59		&38.13		&41.63		&4.38 \\
500	&20.34		&12.86		&33.34		&37.14		&4.36 \\
600	&20.28		&14.31		&29.17		&33.39		&4.21 \\
700	&19.09		&14.49		&26.46		&30.89		&3.81 \\
800	&18.59		&14.15		&24.75		&28.58		&3.63 \\
900	&17.28		&13.59		&23.36		&27.026		&3.34 \\
1000	 &16.27	 	&13.08		&22.07		&25.99		&3.11 \\
2000	 &10.22		&8.12		&16.62		&20.61		&2.17 \\
\hline 
\end{tabular}
\end{table}

\begin{table}[t]
\caption{The calculated cumulative cross sections for production of selected cosmogenic isotopes in \nuc{72}{Ge}. The cross sections are cumulative because they are sums over all isotopes that feed the isotope of interest. }
\label{tab:CrossSections72}
\begin{tabular}{|c|c|c|c|c|c|}

\hline  
Energy            & \multicolumn{5}{c|}{Isotope Production Cross Section (mb)} \\
\cline{2-6}
(MeV)              &\nuc{57}{Co} & \nuc{54}{Mn} &\nuc{68}{Ge}    &\nuc{65}{Zn}    &\nuc{60}{Co}    \\
\hline\hline
0		&0			&0		&0			&0			&0 \\
0		&0			&0		&0			&0			&0 \\
0		&0			&0		&0			&0			&0 \\
0		&0			&0		&0			&0			&0 \\
0		&0			&0		&15.82		&0.15		&0 \\
0		&0			&0		&45.86		&13.43		&0 \\
0		&0			&0		&42.57		&20.87		&0.0010 \\
0		&0			&0		&34.56		&18.01		&0.014 \\
0		&0			&0		&32.2		&19.43		&0.20 \\
0		&0			&0		&30.69		&30.27		&0.61 \\
4.17		&4.32		&0.99	&22.9		&42.50		&5.26 \\
9.45		&10.00		&5.03	&19.88		&38.31		&5.78 \\
12.06	&12.86		&8.55	&16.91		&34.84		&6.22 \\
14.25	&15.12		&11.47	&14.77		&31.37		&6.56 \\
15.13	&16.05		&13.91	&13.46		&28.30		&6.45 \\
14.97	&15.95		&14.48	&12.45		&25.54		&6.09 \\
14.29	&15.22		&14.62	&11.54		&23.63		&5.85 \\
13.46	&14.37		&14.54	&11.13		&21.75		&5.52 \\
12.62	&13.45		&14.07	&10.65		&20.59		&5.14 \\
7.72		&8.24		&8.79	&7.96		&14.69		&3.27 \\
\hline 
\end{tabular}
\end{table} 

\begin{table}[t]
\caption{The calculated cumulative cross sections for production of selected cosmogenic isotopes in \nuc{73}{Ge}. The cross sections are cumulative because they are sums over all isotopes that feed the isotope of interest. }
\label{tab:CrossSections73}
\begin{tabular}{|c|c|c|c|c|c|}

\hline  
Energy            & \multicolumn{5}{c|}{Isotope Production Cross Section (mb)} \\
\cline{2-6}
(MeV)              &\nuc{57}{Co} & \nuc{54}{Mn} &\nuc{68}{Ge}    &\nuc{65}{Zn}    &\nuc{60}{Co}    \\
\hline\hline
\hline 
10		&0			&0		&0			&0			&0 \\
20		&0			&0		&0			&0			&0 \\
30		&0			&0		&0			&0			&0 \\
40		&0			&0		&0			&0			&0 \\
50		&0			&0		&0			&0			&0 \\
60		&0			&0		&4.79		&0.08		&0 \\
70		&0			&0		&22.14		&6.67		&0 \\
80		&0			&0		&23.51		&13.28		&0.002 \\
90		&0			&0		&19.93		&11.61		&0.017 \\
100		&0			&0		&18.37		&12.60		&0.13 \\
200		&3.63		&0.52	&15.22		&35.05		&4.94 \\
300		&7.45		&3.87	&14.00		&33.30		&6.14 \\
400		&10.40		&7.34	&11.98		&30.62		&6.88 \\
500		&12.84		&10.60	&10.39		&27.92		&7.32 \\
600		&14.07		&13.07	&9.20		&25.63		&7.37 \\
700		&13.83		&14.45	&8.41		&23.38		&7.17 \\
800		&13.67		&14.57	&7.84		&21.48		&6.91 \\
900		&13.06		&14.49	&7.25		&19.88		&6.51 \\
1000		&12.51		&13.81	&6.92		&18.58		&6.19 \\
2000		&7.73		&8.64	&5.53		&12.96		&3.94 \\
\hline 
\end{tabular}
\end{table}

\begin{table}[t]
\caption{The calculated cumulative cross sections for production of selected cosmogenic isotopes in \nuc{74}{Ge}. The cross sections are cumulative because they are sums over all isotopes that feed the isotope of interest. }
\label{tab:CrossSections74}
\begin{tabular}{|c|c|c|c|c|c|}

\hline  
Energy            & \multicolumn{5}{c|}{Isotope Production Cross Section (mb)} \\
\cline{2-6}
(MeV)              &\nuc{57}{Co} & \nuc{54}{Mn} &\nuc{68}{Ge}    &\nuc{65}{Zn}    &\nuc{60}{Co}    \\
\hline\hline
10 & 0.00 & 0.00 & 0.00 & 0.00 & 0.00 \\
20 & 0.00 & 0.00 & 0.00 & 0.00 & 0.00 \\
30 & 0.00 & 0.00 & 0.00 & 0.00 & 0.00 \\
40 & 0.00 & 0.00 & 0.00 & 0.00 & 0.00 \\
50 & 0.00 & 0.00 & 0.00 & 0.00 & 0.00 \\
60 & 0.00 & 0.00 & 0.00 & 0.00 & 0.00 \\
70 & 0.00 & 0.00 & 0.57 & 0.00 & 0.00 \\
80 & 0.00 & 0.00 & 7.78 & 1.61 & 0.00 \\
90 & 0.00 & 0.00 & 13.04 & 6.79 & 0.00 \\
100 & 0.00 & 0.00 & 12.79 & 8.06 & 0.01 \\
200 & 1.03 & 0.23 & 10.76 & 28.71 & 4.07 \\
300 & 5.37 & 2.73 & 10.43 & 28.24 & 6.04 \\
400 & 8.16 & 6.14 & 8.74 & 27.13 & 7.28 \\
500 & 10.85 & 9.52 & 7.69 & 25.33 & 8.09 \\
600 & 12.16 & 12.18 & 6.85 & 23.42 & 8.08 \\
700 & 12.77 & 13.66 & 6.17 & 21.17 & 8.19 \\
800 & 12.40 & 14.34 & 5.44 & 19.32 & 7.64 \\
900 & 11.96 & 14.44 & 5.11 & 17.64 & 7.32 \\
1000 & 11.36 & 13.93 & 4.78 & 16.47 & 6.88 \\
2000 & 7.03 & 8.76 & 3.79 & 11.31 & 4.46 \\
\hline 
\end{tabular}
\end{table} 

\begin{table}[t]
\caption{The calculated cumulative cross sections for production of selected cosmogenic isotopes in \nuc{76}{Ge}. The cross sections are cumulative because they are sums over all isotopes that feed the isotope of interest. }
\label{tab:CrossSections76}
\begin{tabular}{|c|c|c|c|c|c|}

\hline  
Energy            & \multicolumn{5}{c|}{Isotope Production Cross Section (mb)} \\
\cline{2-6}
(MeV)              &\nuc{57}{Co} & \nuc{54}{Mn} &\nuc{68}{Ge}    &\nuc{65}{Zn}    &\nuc{60}{Co}    \\
\hline\hline
10 & 0.00 & 0.00 & 0.00 & 0.00 & 0.00 \\
20 & 0.00 & 0.00 & 0.00 & 0.00 & 0.00 \\
30 & 0.00 & 0.00 & 0.00 & 0.00 & 0.00 \\
40 & 0.00 & 0.00 & 0.00 & 0.00 & 0.00 \\
50 & 0.00 & 0.00 & 0.00 & 0.00 & 0.00 \\
60 & 0.00 & 0.00 & 0.00 & 0.00 & 0.00 \\
70 & 0.00 & 0.00 & 0.00 & 0.00 & 0.00 \\
80 & 0.00 & 0.00 & 0.00 & 0.00 & 0.00 \\
90 & 0.00 & 0.00 & 0.04 & 0.00 & 0.00 \\
100 & 0.00 & 0.00 & 1.00 & 0.19 & 0.00 \\
200 & 0.12 & 0.11 & 5.52 & 17.18 & 1.95 \\
300 & 2.43 & 1.24 & 5.92 & 19.91 & 5.07 \\
400 & 4.92 & 3.94 & 5.27 & 20.67 & 7.15 \\
500 & 7.39 & 7.15 & 4.69 & 20.14 & 8.52 \\
600 & 9.16 & 10.34 & 4.11 & 18.97 & 9.12 \\
700 & 9.84 & 12.25 & 3.73 & 17.41 & 9.40 \\
800 & 10.21 & 13.63 & 3.36 & 16.11 & 9.16 \\
900 & 10.02 & 13.86 & 3.08 & 14.84 & 8.66 \\
1000 & 9.62 & 14.00 & 2.85 & 13.60 & 8.36 \\
2000 & 5.82 & 8.92 & 1.82 & 8.67 & 5.43 \\
\hline 
\end{tabular}
\end{table} 

\begin{figure}
\includegraphics[width=8cm]{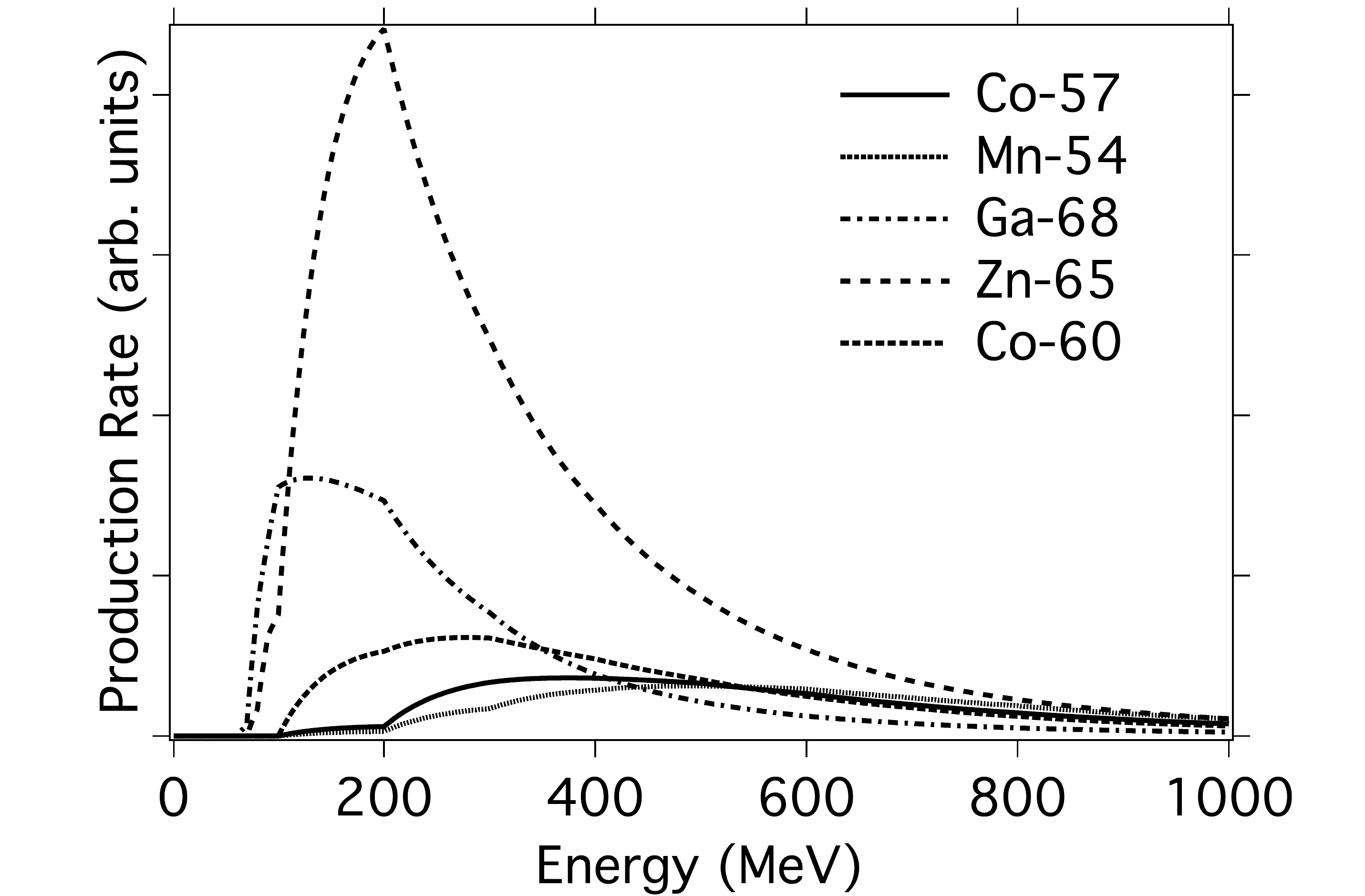}
\caption{The CEM cross sections folded with the cosmic-ray neutron energy spectrum as a function of energy for the isotopes of interest. }
\label{fig:Yield}
\end{figure}

Table~\ref{tab:CrossSectionsExtrap} gives the critical neutron energy range for the majority of the isotope production rate. This low (high) value of this energy range is defined as the energy for which 10\% (90)\% of the cumulative production rate is reached. The ranges are given for both the 4FP60R beam spectrum and the cosmic ray flux. The 4FP60R range also is calculated for the isotopic abundance of our sample, whereas the cosmic ray flux column is calculated for our standard 86\%-14\% isotopic enrichment. Although the energy at which the peak of the production rate for the two spectra are similar, the cosmic ray production rate is still significant at higher energies. Figure~\ref{fig:Yield} shows the production rate as a function of energy for the cosmic ray flux. Since there is a spectral difference between the 4FP60R flux and the cosmic ray flux above 200 MeV, we must use theory along with our measurements to estimate the cosmic-ray production. Therefore, we must consider the possibility that an energy dependence in the theoretical cross section used for this extrapolation will introduce a systematic uncertainty. The input physics to the CEM03.02 cross sections does not change over the energy range of interest and no known physics is omitted. Therefore one does not expect the variation in the cross section uncertainty with energy to be large. References~\cite{gud83,mas05,mas08,Tit08,koro10} indicate that the predictions of the model are uniform across that range. Reference~\cite{Tit08} presents a quantitative analysis of the comparison between the model and previous experiments using a large number of data sets. The result is that the uncertainty is fairly energy independent with a 2-10\% change depending on the data set used for comparison. 

Unfortunately there is a lack of cross section data for large $\Delta$A transitions induced by fast neutrons. Such data would directly validate the theoretical model for interpreting our production rate measurements in terms of the cosmic-ray neutron flux. Ref.~\cite{Tit08}, however, does provide data on proton-induced transitions on an \nuc{56}{Fe} target with $\Delta$A values similar to those of interest in this work. (See Figs. 3-5 in that work.) For each isotope considered here, we use the data of Ref~\cite{Tit08} to estimate the uncertainty that may arise from the energy dependence of the cross section. 

The critical energy range for the cosmic-ray production rate for the most important \BB-background isotope, \nuc{68}{Ge}, is between 110 and 530 MeV. (The low (high) energy of this range was defined by integrating over energy up to the 10\% (90\%) value of the total production rate.) Ref.~\cite{Tit08} studied a few $\Delta$A=8 transitions in \nuc{56}{Fe} (similar to the production of \nuc{68}{Ge} from \nuc{76}{Ge}) that can be used to estimate this uncertainty for the production of \nuc{68}{Ge}. The $\Delta$A=8 transitions in \nuc{56}{Fe} producing nuclei \nuc{48}{Cr}, \nuc{48}{V} and \nuc{48}{Sc} are then used for this comparison. The ratio of the measured and calculated cross sections for these processes varies by an average of $\approx$15\% between 300 and 500 MeV. We estimate the uncertainty in the extrapolation to higher energies as being equal to this 15\% variation scaled to the fraction of model-predicted cosmic-ray production above 200 MeV. The result is given in Table~\ref{tab:CrossSectionsExtrap}.

This uncertainty contribution is different for each isotope due to the different critical neutron energy ranges, the different fraction of cosmic-ray production rate over 200 MeV, and the different results of the comparison of theory to data in Ref.~\cite{Tit08}. For the $\Delta$A=11 production of \nuc{65}{Zn}, we used the data and theory of the isotope production of \nuc{44}{Sc} and \nuc{44}{K} between 300 and 750 MeV with the result that the ratio varies $\approx$35\%. For the $\Delta$A=16 production of \nuc{60}{Co}, we used the data and theory of the isotope production of \nuc{41}{Ar} and \nuc{39}{Cl} between 300 and 750 MeV with the result that the ratio varies  $\approx$50\%. For the $\Delta$A=19 production of \nuc{57}{Co}, we used the data and theory of the isotope production of \nuc{38}{Cl} and \nuc{38}{S} between 300 and 1000 MeV with the result that the ratio varies  $\approx$50\%. For the $\Delta$A=22 production of \nuc{54}{Mn}, we used the data and theory of the isotope production of \nuc{38}{Cl} and \nuc{28}{Mg} between 300 and 1000 MeV with the result that the ratio varies  $\approx$50\%. All these results are summarized in Table~\ref{tab:CrossSectionsExtrap} along with the deduced uncertainty in the estimated cosmic-ray production rate. The uncertainty associated with this scaling dominates the total uncertainty for the cosmic-ray production rate estimates (see Table~\ref{tab:CosmicRateUnc}). Because the production rate of \nuc{68}{Ge} is concentrated at lower energies, the theoretical extrapolation for this important isotope is more reliable than the others.


\begin{table*}[t]
\caption{A summary of the estimates of the uncertainty in the cosmic-ray production rate due to the uncertainty in the energy dependence of the cross section extrapolation. The second column gives the neutron energy range over which the CEM code predicts the majority of the production rate for the 4FP60R beam spectrum and our isotopic sample. The third column gives the neutron energy range for which the CEM code predicts the majority of the production rate for a cosmic ray beam spectrum and a 86\%-14\% enriched sample.The fourth column gives the fraction of the production rate arising from neutrons with energy less than 200 MeV. The fifth column is the estimated uncertainty in the energy dependence of the cross section as described in the text. The last column provides the deduced uncertainty contribution to the cosmic-ray production rate resulting from scaling our measurements.}
\label{tab:CrossSectionsExtrap}
\begin{tabular}{|c|c|c|c|c|c|}

\hline  
Isotope      & Critical E$_n$ for &Critical E$_n$   & Fraction Cosmic Prod.& Est. Uncertainty in Energy  & Deduced Uncertainty in  \\
             & Beam Prod. Rate    & for Cosmic Prod.& for E$_n$ $<$ 200 MeV  & Dependence of Cross Section & Production Rate (\%)    \\
\hline\hline
\nuc{57}{Co} & 170 - 420 MeV      &280 - 1040 MeV           & 1.9\%                & 50\%                             & 49\%   \\
\nuc{54}{Mn} & 200 - 490 MeV      &330 - 1190 MeV           & 1.0\%                & 50\%                             & 50\%   \\
\nuc{68}{Ge} & 35 - 250 MeV       &110 - 530 MeV            & 44.2\%               & 15\%                             & 8.4\%  \\
\nuc{65}{Zn} & 110 - 300 MeV      &140 - 640 MeV            & 29.2\%               & 35\%                             & 24.8\% \\
\nuc{60}{Co} & 140 - 350 MeV      &190 - 890 MeV            & 12.4\%               & 50\%                             & 43.8\%  \\

\hline 
\end{tabular}
\end{table*}

A reader may prefer to use a different cross section model to estimate the production due to cosmic rays. If so, Eqns. \ref{eq:GEANIENFlux} and \ref{eq:flux} can be used with a different model to interpret our measurements of the production rate at WNR in terms of the cosmic ray flux. Furthermore, the uncertainty assigned to the choice of the cross-section model's energy dependence is not based on neutron projectile data, due to a lack of such data. Therefore, one must recognize the caveat this introduces in the extrapolation of the measured production rate values to predicted cosmic-ray production rates. 

\section{Converting the Measured Production Rates to Cosmogenic Production Rates}
To estimate the production rate due to cosmic-ray neutrons, one must know the energy dependent flux of the neutrons and the cross sections.

Ziegler carried out a comprehensive study on the cosmic-ray neutron
flux~\cite{Zieg98} and pointed out that some of the
data from early measurements is incorrect
or of marginal quality. Mei {\em et al.}~\cite{mei09} recognized that previous estimates of the cosmogenic production rates used
various outdated estimates of the cosmic neutron flux.
Improved recent measurements by Gordon {\it et al.}~\cite{Gord04} show that 
the flux density spectrum at sea level can be parameterized as
\begin{eqnarray}
\label{eq:flux}
\phi(E) & = & 1.006\times10^{-6}e^{-0.35\ln^2E+2.1451\ln E}  \nonumber  \\
             & + & 1.011\times10^{-3}e^{-0.4106\ln^2E-0.667\ln E}  
\end{eqnarray}
\noindent
where $E$ is neutron kinetic energy in MeV and $\phi$ is given in units 
of cm$^{-2}$s$^{-1}$MeV$^{-1}$. This parameterization function agrees with the data within $\sim$2\%. Note that parameterization used by Ref.~\cite{ceb06} based on that from Ref.~\cite{Zieg98}
differs from that of Ref.~\cite{Gord04}. Both curves are shown in Fig.~\ref{fig:CosmicNSpectrum}.

\begin{figure}[htb!!!]
\includegraphics[angle=0,width=8.cm] {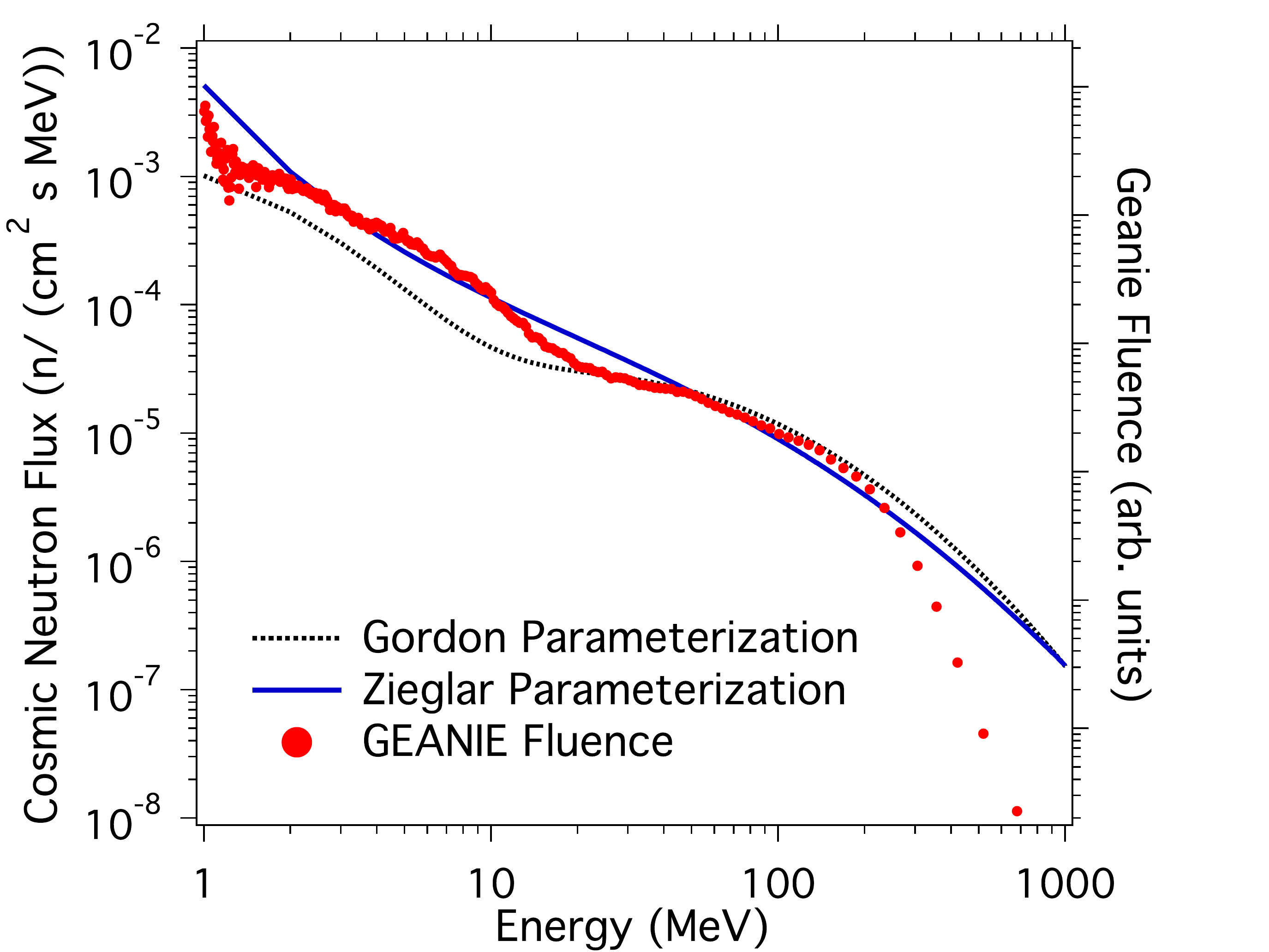}
\caption{\small{The measured neutron flux parameterization functions at sea level~\cite{Gord04,Zieg98} and a normalized plot of the 4FP60R neutron fluence showing that the spectral shape in the critical 100-300 MeV range is similar to the cosmic ray spectrum.}}
\label{fig:CosmicNSpectrum}
\end{figure}

Gordon's measurements show that the shape of the outdoor ground-level neutron spectrum does not depend significantly on altitude, cutoff or solar modulation. Factors depending on atmospheric depth, geomagnetic cutoff, rigidity, and geomagnetic location are given by Gordon for correcting the flux for these effects. His model is slightly different than Ziegler or the JEDEC standard. As a comparison, the cosmic production rates resulting from Gordon's flux are 13-33\% higher than that using the flux of Ziegler for the isotopes of interest in this work.

The ratio of $N_0$ to $N^{Pred}_0$ values given in Table~\ref{tab:CountResults} can be used to scale a predicted production rate ($K_{Gordon}$) based on the cross sections of Section~\ref{sec:CrossSection} and the Gordon neutron flux parameterization to provide an estimate ($K_{scaled}$) of the cosmogenic production rate indicated by our measurements. The numbers used in this arithmetic are summarized in Table~\ref{tab:CosmicRates} with the results also quoted in Table~\ref{tab:EnrGeProdRate} for comparison to previous predictions.

\begin{table}[t]

\caption{A summary of the parameters that enter into the calculation of our estimate for the cosmic-ray production rate of the various isotopes ($K_{scaled}$). The uncertainties quoted for $K_{scaled}$ are explained in Table~\ref{tab:CosmicRateUnc}.  The value for \nuc{57}{Co} is taken as that of the 122-kev line as it has a much smaller total uncertainty. The \nuc{60}{Co} value is the average of the two individual measurements and, since the uncertainty is dominated by systematic effects, we quote the larger of the two associated total uncertainty values given in Table~\ref{tab:CosmicRateUnc}. These production rate values are for a nominal 86\% \nuc{76}{Ge} 14\% \nuc{74}{Ge} isotopic mixture.}
\label{tab:CosmicRates}
\begin{tabular}{|c|c|c|c|}

\hline  
Isotope		& Ratio 						& $K_{Gordon}$  & $K_{scaled}$  \\
Isotope		& ($\frac{N_0}{N^{Pred}_0})$	& (atoms/(kg d) & (atoms/(kg d)  \\
\hline\hline
\nuc{57}{Co} &  $0.25 \pm 0.01$		&   2.93		& $0.72 \pm 0.37$                              \\
\nuc{54}{Mn}	&   $0.67 \pm 0.03$		&   2.91		& $1.96\pm 1.01$                            \\
\nuc{68}{Ge}	&   $0.36 \pm 0.03$		&   5.83		& $2.12\pm 0.39$                              \\
\nuc{65}{Zn}	&   $0.55 \pm 0.01$		&   16.24	& $8.94\pm 2.53$                              \\
\nuc{60}{Co}	&   $0.63 \pm 0.03$		&   4.06		& $2.55\pm 1.20$                              \\
\hline 
\end{tabular}
\end{table} 

\begin{table*}[t]

\caption{A summary of the uncertainties (in \%) that contribute to the total uncertainty of cosmic production rate. The column labeled {\em SubTotal}
refers to the quadrature sum of all non-cosmic-ray related contributions and provides the uncertainty for the ratio in Table~\ref{tab:CosmicRates}.}
\label{tab:CosmicRateUnc}
\begin{tabular}{|c|c|c|c|c|c||c||c|c|c||c||}

\hline  
Isotope				& Counting 		&	Efficiency &  Source    & Predicted	&Flux			& SubTotal	& Cosmic	 	&Neutron		&Proton        &Total \\
(Line Energy)		& Statistics		&			   &  Activity  & 4FP60R 	 & Chamber      &           &  Neutron 	& Specral    &Correction    &\\
              		&				&			   &              & Fluence    &Live Time      &           &  Flux      &Difference  &    &\\
\hline\hline
\nuc{57}{Co}(122)		&2.9			&0.4	        &1.0        &0.9             &0.3            &3.2       		 &12.5  	& 49        &5.0        	&51 \\
\nuc{57}{Co}(136)		&16.3		&1.2			&1.0			&0.9				&0.3				&16.4			&12.5	 & 49		&5.0				&53\\
\nuc{54}{Mn}				&4.0			&1.0			&1.0			&1.2				&0.3				&4.4				&12.5	& 50		&5.0				&51\\
\nuc{68}{Ge}				&9.1			&2.4			&1.0			&0.5				&0.3				&9.5				&12.5	&8.4	      	&5.0				&18\\
\nuc{65}{Zn}				&1.1			&2.2			&1.0			&0.5				&0.3				&2.7				&12.5	&24.8			&5.0				&28\\
\nuc{60}{Co}(1173)		&3.0			&3.0			&1.0			&0.6				&0.3				&4.5				&12.5	&44			&5.0				&46\\
\nuc{60}{Co}(1332)		&3.2 		&1.8			&1.0			&0.6				&0.3				&3.9				&12.5	&44			&5.0				&46\\
\hline
\end{tabular}
\end{table*}

The estimate of the cosmogenic production from these measured results have some additional systematic uncertainties. These include the precision to which the cosmic ray neutron flux is known (10-15\% and so we split the difference and use 12.5\% \cite{Gord04}). The total cosmogenic rate includes contributions from subdominant proton and pion interactions. These only contribute approximately 10\% \cite{bara06} to the total production rate. These charged particles are much less penetrating than neutrons and therefore their impact on any given sample is very geometry dependent. Hence we assume a 50\% uncertainly on this correction for an uncertainty of 5\%. Again these uncertainties are uncorrelated and result in a estimated systematic uncertainty of 13.5\%. 

We estimated the rate of cosmogenic production in \nuc{enr}{Ge} with isotopic abundance limited to isotopes 74 (14\%) and 76 (86\%). However, a small admixture of isotope 70 can result in significantly more \nuc{68}{Ge} due to the much lower threshold for the neutron reaction and the higher cross section. Here we present a rough formula to estimate the \nuc{68}{Ge} production rate ($P_{68}$) as a function of the amount of \nuc{70}{Ge} present in the sample in per cent ($X$).

\begin{equation}
\label{eq:Ge68Rate}
P_{68}  = (2.12 + 0.75X) {\mbox atoms/(kg~d)}
\end{equation}

This formula should be accurate to approximately 20\%. The other critical isotope for \BB\ research is \nuc{60}{Co} and it has a small production dependence on the fraction of \nuc{70}{Ge}.

\section{Discussion and Conclusion}

We measured the production of \nuc{57}{Co}, \nuc{54}{Mn}, \nuc{68}{Ge}, \nuc{65}{Zn}, and \nuc{60}{Co} in a sample of Ge enriched in isotope 76 due to high-energy neutron interactions within a neutron beam with a spectrum similar to that of the cosmic-ray neutron flux at the Earth's surface. The uncertainty on the beam-produced production rate at those energies is below 10\%, depending on isotope, as given in Table~\ref{tab:CosmicRateUnc}. The results, presented in Table~\ref{tab:CountResults}, were compared to cross sections calculated with CEM03.02. The measurements are smaller by about a factor of 2-4 than these calculations depending on isotope.  We scaled these measurements using CEM03.02 cross sections to estimate the cosmic-ray production rate of these troublesome isotopes and present the results in Table~\ref{tab:CosmicRates}. Our estimated total uncertainty for the production rate of the critical \nuc{68}{Ge} isotope is approximately 20\% providing much better guidance to double-beta decay experimenters in their efforts to understand background due to this isotope. The uncertainties on the deduced cosmogenic production is dominated by the uncertainty in the extrapolation of the cross sections to higher neutron energies. This large uncertainty arises due to a lack of large $\Delta$A, neutron-induced reactions at higher incident energies with which to validate the theory used for the extrapolation.

\begin{acknowledgments}
 We gratefully acknowledge the support of the U.S. Department of Energy through
the LANL/LDRD Program for this work. We thank Frank Avignone III for providing the enriched Ge sample and we thank Jason Detwiler for a careful reading of this manuscript. 
 This work benefited from the use of the Los Alamos Neutron Science Center, funded
 by the U.S. Department of Energy under contract DE-AC52-06NA25396. 
We are grateful for the ToF SIMS measurements that were performed  by Zihua Zhu using EMSL, a national scientific user facility sponsored by the Department of Energy's Office of Biological and Environmental Research and located at Pacific Northwest National Laboratory. We thank Richard Kouzes for making arrangements for the ToF SIMS measurements. 
 This work also benefited from our underground laboratory at the Waste Isolation Pilot Plant (WIPP), which we operate with support from the Nuclear Physics office of the U.S. Department of Energy under contract number 2011LANLE9BW. Finally, we thank our friends and hosts at the Waste Isolation Pilot Plant (WIPP) for their continuing support of our activities underground at that facility.  
\end{acknowledgments}


\section*{References}

\bibliography{DoubleBetaDecay.bbl}

\begin{thebibliography}{34}
\expandafter\ifx\csname natexlab\endcsname\relax\def\natexlab#1{#1}\fi
\expandafter\ifx\csname bibnamefont\endcsname\relax
  \def\bibnamefont#1{#1}\fi
\expandafter\ifx\csname bibfnamefont\endcsname\relax
  \def\bibfnamefont#1{#1}\fi
\expandafter\ifx\csname citenamefont\endcsname\relax
  \def\citenamefont#1{#1}\fi
\expandafter\ifx\csname url\endcsname\relax
  \def\url#1{\texttt{#1}}\fi
\expandafter\ifx\csname urlprefix\endcsname\relax\def\urlprefix{URL }\fi
\providecommand{\bibinfo}[2]{#2}
\providecommand{\eprint}[2][]{\url{#2}}

\bibitem[{\citenamefont{Elliott and Vogel}(2002)}]{Ell02}
\bibinfo{author}{\bibfnamefont{S.~R.} \bibnamefont{Elliott}} \bibnamefont{and}
  \bibinfo{author}{\bibfnamefont{P.}~\bibnamefont{Vogel}},
  \bibinfo{journal}{Ann.\ Rev.\ Nucl.\ Part.\ Sci.}
  \textbf{\bibinfo{volume}{52}}, \bibinfo{pages}{115} (\bibinfo{year}{2002}).

\bibitem[{\citenamefont{Elliott and Engel}(2004)}]{Ell04}
\bibinfo{author}{\bibfnamefont{S.~R.} \bibnamefont{Elliott}} \bibnamefont{and}
  \bibinfo{author}{\bibfnamefont{J.}~\bibnamefont{Engel}},
  \bibinfo{journal}{J.\ Phys.\ G:\ Nucl.\ Part.\ Phys.}
  \textbf{\bibinfo{volume}{30}}, \bibinfo{pages}{R 183} (\bibinfo{year}{2004}).

\bibitem[{\citenamefont{Barabash et~al.}(2004)\citenamefont{Barabash, Hubert,
  Huber, and Umatov}}]{Bar04}
\bibinfo{author}{\bibfnamefont{A.~S.} \bibnamefont{Barabash}},
  \bibinfo{author}{\bibfnamefont{F.}~\bibnamefont{Hubert}},
  \bibinfo{author}{\bibfnamefont{P.}~\bibnamefont{Huber}}, \bibnamefont{and}
  \bibinfo{author}{\bibfnamefont{V.~I.} \bibnamefont{Umatov}},
  \bibinfo{journal}{Phys. At. Nucl.} \textbf{\bibinfo{volume}{67}},
  \bibinfo{pages}{438} (\bibinfo{year}{2004}).

\bibitem[{\citenamefont{Avignone et~al.}(2005)\citenamefont{Avignone, King, and
  Zdesenko}}]{Avi05}
\bibinfo{author}{\bibfnamefont{F.~T.~{\protect III}.} \bibnamefont{Avignone}},
  \bibinfo{author}{\bibfnamefont{G.~S.~{\protect III}.} \bibnamefont{King}},
  \bibnamefont{and} \bibinfo{author}{\bibfnamefont{Y.}~\bibnamefont{Zdesenko}},
  \bibinfo{journal}{New Journal of Physics} \textbf{\bibinfo{volume}{7}},
  \bibinfo{pages}{6} (\bibinfo{year}{2005}).

\bibitem[{\citenamefont{Ejiri}(2005)}]{eji05}
\bibinfo{author}{\bibfnamefont{H.}~\bibnamefont{Ejiri}}, \bibinfo{journal}{J.
  Phys. Soc. Jap.} \textbf{\bibinfo{volume}{74}}, \bibinfo{pages}{2101}
  (\bibinfo{year}{2005}).

\bibitem[{\citenamefont{Avignone et~al.}(2008)\citenamefont{Avignone, Elliott,
  and Engel}}]{avi08}
\bibinfo{author}{\bibfnamefont{F.~T.~{\protect III}.} \bibnamefont{Avignone}},
  \bibinfo{author}{\bibfnamefont{S.~R.} \bibnamefont{Elliott}},
  \bibnamefont{and} \bibinfo{author}{\bibfnamefont{J.}~\bibnamefont{Engel}},
  \bibinfo{journal}{Rev. Mod. Phys.} \textbf{\bibinfo{volume}{80}},
  \bibinfo{pages}{481} (\bibinfo{year}{2008}), \eprint{arXiv:0708.1033}.

\bibitem[{\citenamefont{Aalseth et~al.}(2002)}]{aal02a}
\bibinfo{author}{\bibfnamefont{C.~E.} \bibnamefont{Aalseth}}
  \bibnamefont{et~al.} (\bibinfo{collaboration}{IGEX}),
  \bibinfo{journal}{Phys.\ Rev. D.} \textbf{\bibinfo{volume}{65}},
  \bibinfo{pages}{092007} (\bibinfo{year}{2002}).

\bibitem[{\citenamefont{Baudis et~al.}(1999)}]{bau99}
\bibinfo{author}{\bibfnamefont{L.}~\bibnamefont{Baudis}} \bibnamefont{et~al.},
  \bibinfo{journal}{Phys.\ Rev.\ Lett.} \textbf{\bibinfo{volume}{83}},
  \bibinfo{pages}{41} (\bibinfo{year}{1999}).

\bibitem[{\citenamefont{Klapdor-Kleingrothaus and Krivosheina}(2006)}]{kla06}
\bibinfo{author}{\bibfnamefont{H.~V.} \bibnamefont{Klapdor-Kleingrothaus}}
  \bibnamefont{and} \bibinfo{author}{\bibfnamefont{I.~V.}
  \bibnamefont{Krivosheina}}, \bibinfo{journal}{Mod. Phys. Lett. A}
  \textbf{\bibinfo{volume}{21}}, \bibinfo{pages}{1547} (\bibinfo{year}{2006}).

\bibitem[{\citenamefont{Elliott et~al.}(2010)}]{ell08}
\bibinfo{author}{\bibfnamefont{S.}~\bibnamefont{Elliott}} \bibnamefont{et~al.},
  \emph{\bibinfo{title}{Proceedings of the Carolina International Symposium on
  Neutrino Physics}}, vol. \bibinfo{volume}{173} (\bibinfo{publisher}{IOP
  Publishing}, \bibinfo{address}{London}, \bibinfo{year}{2010}),
  \eprint{arXiv:0807.1741}.

\bibitem[{\citenamefont{Guiseppe et~al.}(2008)}]{gui08}
\bibinfo{author}{\bibfnamefont{V.~E.} \bibnamefont{Guiseppe}}
  \bibnamefont{et~al.}, \bibinfo{journal}{Nucl. Sci. Symp. Conf. Rec. NSS'08}
  p. \bibinfo{pages}{1793} (\bibinfo{year}{2008}), \eprint{arXiv:0811.2446}.

\bibitem[{\citenamefont{Henning et~al.}(2009)}]{hen09}
\bibinfo{author}{\bibfnamefont{R.}~\bibnamefont{Henning}} \bibnamefont{et~al.}
  (\bibinfo{year}{2009}), \eprint{arXiv:0907.1581}.

\bibitem[{\citenamefont{Sch{\protect\"{o}}nert et~al.}(2005)}]{sch05}
\bibinfo{author}{\bibfnamefont{S.}~\bibnamefont{Sch{\protect\"{o}}nert}}
  \bibnamefont{et~al.}, \bibinfo{journal}{Nucl. Phys. Proc. Suppl.}
  \textbf{\bibinfo{volume}{145}}, \bibinfo{pages}{242} (\bibinfo{year}{2005}).

\bibitem[{\citenamefont{Miley et~al.}(1992)\citenamefont{Miley, Avignone,
  Brodzinski, Hensley, and Reeves}}]{mil92}
\bibinfo{author}{\bibfnamefont{H.~S.} \bibnamefont{Miley}},
  \bibinfo{author}{\bibfnamefont{F.~T.} \bibnamefont{Avignone}},
  \bibinfo{author}{\bibfnamefont{R.~L.} \bibnamefont{Brodzinski}},
  \bibinfo{author}{\bibfnamefont{W.~K.} \bibnamefont{Hensley}},
  \bibnamefont{and} \bibinfo{author}{\bibfnamefont{J.~H.}
  \bibnamefont{Reeves}}, \bibinfo{journal}{Nucl. Phys. B (Proc. Suppl.)}
  \textbf{\bibinfo{volume}{28A}}, \bibinfo{pages}{212} (\bibinfo{year}{1992}).

\bibitem[{\citenamefont{Avignone et~al.}(1992)}]{avi92}
\bibinfo{author}{\bibfnamefont{F.~T.~{\protect III}.} \bibnamefont{Avignone}}
  \bibnamefont{et~al.}, \bibinfo{journal}{Nucl. Phys. B (Proc. Suppl)}
  \textbf{\bibinfo{volume}{28A}}, \bibinfo{pages}{280} (\bibinfo{year}{1992}).

\bibitem[{\citenamefont{Cebri\'{a}n et~al.}(2006)}]{ceb06}
\bibinfo{author}{\bibfnamefont{S.}~\bibnamefont{Cebri\'{a}n}}
  \bibnamefont{et~al.}, \bibinfo{journal}{Journal of Physics: Conference
  Series} \textbf{\bibinfo{volume}{39}}, \bibinfo{pages}{344}
  (\bibinfo{year}{2006}), \bibinfo{note}{{T}AUP 2005: Proc. Ninth Int. Conf. on
  Topics in Astroparticle and Underground Physics}.

\bibitem[{\citenamefont{Horiguchi et~al.}(1983)}]{hor83}
\bibinfo{author}{\bibfnamefont{T.}~\bibnamefont{Horiguchi}}
  \bibnamefont{et~al.}, \bibinfo{journal}{Int. J. Appl. Rad. Isot.}
  \textbf{\bibinfo{volume}{34}}, \bibinfo{pages}{1531} (\bibinfo{year}{1983}).

\bibitem[{\citenamefont{Aleksandrov et~al.}(1995)}]{ale95}
\bibinfo{author}{\bibfnamefont{Y.~V.} \bibnamefont{Aleksandrov}}
  \bibnamefont{et~al.}, \bibinfo{journal}{Bull. Russ. Acad. Sci. - Phys. Ser.}
  \textbf{\bibinfo{volume}{59}}, \bibinfo{pages}{895} (\bibinfo{year}{1995}).

\bibitem[{\citenamefont{Norman et~al.}(2005)}]{nor05}
\bibinfo{author}{\bibfnamefont{E.~B.} \bibnamefont{Norman}}
  \bibnamefont{et~al.}, \bibinfo{journal}{Nucl. Phys. B (Proc. Suppl.)}
  \textbf{\bibinfo{volume}{143}}, \bibinfo{pages}{508} (\bibinfo{year}{2005}).

\bibitem[{\citenamefont{Barabanov et~al.}(2006)\citenamefont{Barabanov,
  Belogurov, Bezrukov, Denisov, Kornoukhov, and Sobolevsky}}]{bara06}
\bibinfo{author}{\bibfnamefont{I.}~\bibnamefont{Barabanov}},
  \bibinfo{author}{\bibfnamefont{S.}~\bibnamefont{Belogurov}},
  \bibinfo{author}{\bibfnamefont{L.}~\bibnamefont{Bezrukov}},
  \bibinfo{author}{\bibfnamefont{A.}~\bibnamefont{Denisov}},
  \bibinfo{author}{\bibfnamefont{V.}~\bibnamefont{Kornoukhov}},
  \bibnamefont{and}
  \bibinfo{author}{\bibfnamefont{N.}~\bibnamefont{Sobolevsky}},
  \bibinfo{journal}{Nucl. Instrum. Meth.B} \textbf{\bibinfo{volume}{251}},
  \bibinfo{pages}{115Ð120} (\bibinfo{year}{2006}).

\bibitem[{\citenamefont{Mei et~al.}(2009)\citenamefont{Mei, Yin, and
  Elliott}}]{mei09}
\bibinfo{author}{\bibfnamefont{D.-M.} \bibnamefont{Mei}},
  \bibinfo{author}{\bibfnamefont{Z.-B.} \bibnamefont{Yin}}, \bibnamefont{and}
  \bibinfo{author}{\bibfnamefont{S.~R.} \bibnamefont{Elliott}},
  \bibinfo{journal}{Astropart. Phys.} \textbf{\bibinfo{volume}{31}},
  \bibinfo{pages}{417Ð420} (\bibinfo{year}{2009}), \eprint{arXiv:0903.2273}.

\bibitem[{\citenamefont{Balysh et~al.}(1992)}]{bal92}
\bibinfo{author}{\bibfnamefont{A.}~\bibnamefont{Balysh}} \bibnamefont{et~al.},
  in \emph{\bibinfo{booktitle}{Proceedings of the XXVIIth Rencontre de Moriond
  Progress in Atomic Physics Neutrinos and Gravitation}}
  (\bibinfo{publisher}{Editions Frontieres}, \bibinfo{address}{Singapore},
  \bibinfo{year}{1992}), p. \bibinfo{pages}{177}.

\bibitem[{\citenamefont{Back and Ramachers}(2008)}]{bac08}
\bibinfo{author}{\bibfnamefont{J.~J.} \bibnamefont{Back}} \bibnamefont{and}
  \bibinfo{author}{\bibfnamefont{Y.~A.} \bibnamefont{Ramachers}},
  \bibinfo{journal}{Nucl. Instrum. Meth. A} \textbf{\bibinfo{volume}{586}},
  \bibinfo{pages}{286} (\bibinfo{year}{2008}).

\bibitem[{\citenamefont{Koning et~al.}(2004)\citenamefont{Koning, Hilaire, and
  Duijvestijn}}]{kon04}
\bibinfo{author}{\bibfnamefont{A.~J.} \bibnamefont{Koning}},
  \bibinfo{author}{\bibfnamefont{S.}~\bibnamefont{Hilaire}}, \bibnamefont{and}
  \bibinfo{author}{\bibfnamefont{M.~C.} \bibnamefont{Duijvestijn}}, in
  \emph{\bibinfo{booktitle}{Proceedings of the International Conference on
  Nuclear Data for Science and Technology - ND2004}}, edited by
  \bibinfo{editor}{\bibfnamefont{R.~C.} \bibnamefont{Haight}},
  \bibinfo{editor}{\bibfnamefont{M.~B.} \bibnamefont{Chadwick}},
  \bibinfo{editor}{\bibfnamefont{T.}~\bibnamefont{Kawano}}, \bibnamefont{and}
  \bibinfo{editor}{\bibfnamefont{P.}~\bibnamefont{Talou}}
  (\bibinfo{year}{2004}), vol. \bibinfo{volume}{769}, p. \bibinfo{pages}{1154}.

\bibitem[{\citenamefont{Ziegler}(1998)}]{Zieg98}
\bibinfo{author}{\bibfnamefont{J.~F.} \bibnamefont{Ziegler}},
  \bibinfo{journal}{IBM J. Res. and Develop.} \textbf{\bibinfo{volume}{42}},
  \bibinfo{pages}{117} (\bibinfo{year}{1998}).

\bibitem[{\citenamefont{Gordon et~al.}(2004)\citenamefont{Gordon, Goldhagen
  et~al.}}]{Gord04}
\bibinfo{author}{\bibfnamefont{M.~S.} \bibnamefont{Gordon}},
  \bibinfo{author}{\bibfnamefont{P.}~\bibnamefont{Goldhagen}},
  \bibnamefont{et~al.}, \bibinfo{journal}{IEEE Transactions on Nuclear Science}
  \textbf{\bibinfo{volume}{51}}, \bibinfo{pages}{3427} (\bibinfo{year}{2004}).

\bibitem[{\citenamefont{Klapdor-Kleingrothaus et~al.}(2002)}]{kla02b}
\bibinfo{author}{\bibfnamefont{H.~V.} \bibnamefont{Klapdor-Kleingrothaus}}
  \bibnamefont{et~al.}, \bibinfo{journal}{Nucl. Instrum. Meth. A}
  \textbf{\bibinfo{volume}{481}}, \bibinfo{pages}{149} (\bibinfo{year}{2002}).

\bibitem[{\citenamefont{Lisowski et~al.}(1990)\citenamefont{Lisowski, Bowman,
  Russell, and Wender}}]{lis90}
\bibinfo{author}{\bibfnamefont{P.~W.} \bibnamefont{Lisowski}},
  \bibinfo{author}{\bibfnamefont{C.~D.} \bibnamefont{Bowman}},
  \bibinfo{author}{\bibfnamefont{G.~J.} \bibnamefont{Russell}},
  \bibnamefont{and} \bibinfo{author}{\bibfnamefont{S.~A.}
  \bibnamefont{Wender}}, \bibinfo{journal}{Nucl. Sci. Eng.}
  \textbf{\bibinfo{volume}{106}}, \bibinfo{pages}{208} (\bibinfo{year}{1990}).

\bibitem[{\citenamefont{Becker and Nelson}(1997)}]{bec97}
\bibinfo{author}{\bibfnamefont{J.~A.} \bibnamefont{Becker}} \bibnamefont{and}
  \bibinfo{author}{\bibfnamefont{R.~O.} \bibnamefont{Nelson}},
  \bibinfo{journal}{Nucl. Phys. News} \textbf{\bibinfo{volume}{7}},
  \bibinfo{pages}{11} (\bibinfo{year}{1997}).

\bibitem[{\citenamefont{Gudima et~al.}(1983)\citenamefont{Gudima, Mashnik, and
  Toneev}}]{gud83}
\bibinfo{author}{\bibfnamefont{K.~K.} \bibnamefont{Gudima}},
  \bibinfo{author}{\bibfnamefont{S.~G.} \bibnamefont{Mashnik}},
  \bibnamefont{and} \bibinfo{author}{\bibfnamefont{V.~D.}
  \bibnamefont{Toneev}}, \bibinfo{journal}{Nucl. Phys. A}
  \textbf{\bibinfo{volume}{401}}, \bibinfo{pages}{329} (\bibinfo{year}{1983}).

\bibitem[{\citenamefont{Mashnik et~al.}(2005)\citenamefont{Mashnik, Baznat,
  Gudima, Sierk, and Prael}}]{mas05}
\bibinfo{author}{\bibfnamefont{S.~G.} \bibnamefont{Mashnik}},
  \bibinfo{author}{\bibfnamefont{M.~I.} \bibnamefont{Baznat}},
  \bibinfo{author}{\bibfnamefont{K.~K.} \bibnamefont{Gudima}},
  \bibinfo{author}{\bibfnamefont{A.~J.} \bibnamefont{Sierk}}, \bibnamefont{and}
  \bibinfo{author}{\bibfnamefont{R.~E.} \bibnamefont{Prael}},
  \bibinfo{journal}{J. Nucl. Radiochem. Sci.} \textbf{\bibinfo{volume}{6}},
  \bibinfo{pages}{A1} (\bibinfo{year}{2005}), \eprint{nucl-th/0503061}.

\bibitem[{\citenamefont{Mashnik et~al.}(2008)\citenamefont{Mashnik, Gudima,
  Prael, Sierk, Baznat, and Mokhov}}]{mas08}
\bibinfo{author}{\bibfnamefont{S.~G.} \bibnamefont{Mashnik}},
  \bibinfo{author}{\bibfnamefont{K.~K.} \bibnamefont{Gudima}},
  \bibinfo{author}{\bibfnamefont{R.~E.} \bibnamefont{Prael}},
  \bibinfo{author}{\bibfnamefont{A.~J.} \bibnamefont{Sierk}},
  \bibinfo{author}{\bibfnamefont{M.~I.} \bibnamefont{Baznat}},
  \bibnamefont{and} \bibinfo{author}{\bibfnamefont{N.~V.}
  \bibnamefont{Mokhov}}, in \emph{\bibinfo{booktitle}{Invited lectures
  presented at the Joint ICTP-IAEA Advanced Workshop on Model Codes for
  Spallation Reactions}} (\bibinfo{year}{2008}), p.~\bibinfo{pages}{51},
  \bibinfo{note}{{L}A-UR-08-2931, Los Alamos (2008); IAEA Report
  INDC(NDS)-0530, Distr. SC, Vienna, Austria, August 2008},
  \eprint{arXiv:0805.0751v2 [nucl-th]}.

\bibitem[{\citenamefont{\protect{ \protect{Yu}}. E.~Titarenko
  et~al.}(2008)}]{Tit08}
\bibinfo{author}{\bibnamefont{\protect{ \protect{Yu}}. E.~Titarenko}}
  \bibnamefont{et~al.}, \bibinfo{journal}{Phys. Rev. C}
  \textbf{\bibinfo{volume}{78}}, \bibinfo{pages}{034615}
  (\bibinfo{year}{2008}).

\bibitem[{\citenamefont{Korovin et~al.}(2010)\citenamefont{Korovin, Natalenko,
  Konobeyev, Stankovskiy, and Mashnik}}]{koro10}
\bibinfo{author}{\bibfnamefont{Y.~A.} \bibnamefont{Korovin}},
  \bibinfo{author}{\bibfnamefont{A.~A.} \bibnamefont{Natalenko}},
  \bibinfo{author}{\bibfnamefont{A.~Y.} \bibnamefont{Konobeyev}},
  \bibinfo{author}{\bibfnamefont{A.~Y.} \bibnamefont{Stankovskiy}},
  \bibnamefont{and} \bibinfo{author}{\bibfnamefont{S.~G.}
  \bibnamefont{Mashnik}} (\bibinfo{year}{2010}), \bibinfo{note}{submitted to
  NIM A}, \eprint{arXiv:1003.2225}.

\end{thebibliography}

\end{document}